\documentclass[aps,pre,showpacs,twocolumn]{revtex4}
\usepackage{float,graphicx}
\usepackage{psfig}

\usepackage{dcolumn}
\newcommand{\ti}{\textit}
\newcommand{\be}{\begin{equation}}
\newcommand{\ee}{\end{equation}}
\newcommand{\bea}{\begin{eqnarray}}
\newcommand{\eea}{\end{eqnarray}}
\newcommand{\la}{\langle}
\newcommand{\rar}{\rightarrow}
\newcommand{\ra}{\rangle}

\begin{document}

\title{Ring structures and mean first passage time in networks}

\author{Andrea Baronchelli and Vittorio Loreto}

\affiliation{INFM and Dipartimento di Fisica, Universit\`a di Roma
``La Sapienza'' and INFM Center for Statistical Mechanics and
Complexity (SMC),\\ Piazzale A. Moro 2, 00185 Roma, Italy}

\date{\today}

\begin{abstract}
In this paper we address the problem of the calculation of the mean
first passage time (MFPT) on generic graphs. We focus in particular on
the mean first passage time on a node $s$ for a random walker starting
from a generic, unknown, node $x$. We introduce an approximate scheme
of calculation which maps the original process in a Markov process in
the space of the so-called {\em rings}, described by a transition
matrix of size $O(\ln N / \ln \la k \ra \times \ln N / \ln \la k
\ra)$, where $N$ is the size of the graph and $\la k \ra$ the average
degree in the graph. In this way one has a drastic reduction of
degrees of freedom with respect to the size $N$ of the transition
matrix of the original process, corresponding to an extremely-low
computational cost. We first apply the method to the Erd\"os-Renyi
random graph for which the method allows for almost perfect agreement
with numerical simulations. Then we extend the approach to the
Barabasi-Albert graph, as an example of scale-free graph, for which
one obtains excellent results. Finally we test the method with two
real world graphs, Internet and a network of the brain, for which we
obtain accurate results.
\end{abstract}
\maketitle

\section{Introduction}

Modern graph theory starts with the study of Erd\"os-Renyi random
networks in 1959~\cite{ER}. In more recent times it has regained a
great amount of attention~\cite{bollobas} since it has become evident
that many different system can be described as complex scale free
networks i.e assemblies of nodes and edges with nontrivial topological
properties~\cite{Barabasi02, Newman03}.  \\ In this article we focus
on the properties of random walks on generic graphs. It is well known
that random walk is a fundamental process to explore an
environment~\cite{Hughes,fpp,L,AH00,AF,Woess}, and recently great
attention has been devoted to the study of random walk on networks
(see, for instance, \cite{Adamic, Almaas, newmanbetw, nohrieger, Gallos, Sood,
Bollt, Saramaki, Evans, yang}).  In this process a walker, situated on a
given node at time $t$, can be found with probability $1/k$ on any of
the $k$ neighbors of that node at time $t+1$.  \\ In particular we are
interested in the mean first passage time (MFPT) on a node $s$ for a
random walker starting from a generic, unknown, node $x$. It is
important to note here that Noh and Rieger~\cite{nohrieger} have
derived, exploiting the properties of Markov chains, an exact formula
for the MFPT $ T_{sj} $ of a random walker between two nodes $s$ and
$j$ in a generic finite network.  \\ In this paper, however, we do not
trivially average $T_{sj} $ over all $j \neq s$, a very costly
operation, but we use the concept of \ti{ring} (see
also~\cite{costa}). In this perspective we study the graph as seen by
node $s$, and partition it in rings according to the topological
distance of the different nodes from $s$ (see
also~\cite{egocentered,Sood}).  This allows us to map the original
Markov problem (of $N$ states) in a new Markov chain of drastically
reduced dimension ($O(\ln N / \ln \la k \ra \times \ln N / \ln \la
k \ra)$) and, as a consequence, to calculate MFPT on a generic node
$s$ with a reduced computational cost. On the other hand with the new
process the identity of the single target node $s$ is lost, and all
the nodes with the same connectivity (i.e. number of neighbors) are
not distinguishable.  \\ Our explicit calculation is almost free of
approximations only for Erd\"os-Renyi random graphs, for which we
obtain an excellent agreement between theory and numerical
simulations. The more disordered scenario of other complex networks
makes the extension of our approach progressively more
problematic. Nevertheless we find quite surprisingly that our approach
is able to make very good predictions also for other synthetic
networks, such as the Barabasi-Albert scale free
networks~\cite{barabasiI}, and at least two real world graphs. In all
these cases, the considered networks behave, with respect to the
property studied, as if they were random graph with the same average
degree.  Finally our approach allows us to show that a random walker
recovers rapidly the degree distribution of the network it is
exploring.  \\ The paper is organized as follows. In section II the
concept of ring is introduced and the new Markov process on which the
original problem can be mapped is defined. In section III explicit
calculations for the case of random graphs are performed. It is shown
that the description of a random graph in term of rings is very
accurate, and that theoretical predictions for the MFPT are in
excellent agreement with results from simulations. Section IV,
finally, is devoted to the possible extension of the theory to other
networks. Notwithstanding the difficulties that arise in the
analytical extension of the theory, it is shown that MFPT of walkers
in both artificial and real life networks can be predicted quite
accurately with our theory. It is also shown that a random walk can be
used for the reconstruction of the degree distribution of a network.

\section{The new process - rings}

All the information about a generic graph is contained in the
adjacency matrix $A$ whose element $A_{ij}=1$ if nodes $i$ and $j$ are
connected, and $A_{ij}=0$ otherwise. We shall consider here only
undirected graphs, i.e. $A_{ij}=A_{ji}$, which do not contain any
links connecting a node with itself ($A_{ii}=0, \; \forall i$).  The degree of a
node, $k(i)$, is given by $k(i)=\sum_j A_{ij}$.  Finally, we shall
concentrate only on the case of connected graphs, i.e. graphs in which
each pair of nodes $i,j$, with $k(i),k(j) \neq 0$, are connected with
at least one single path.  From a random walk point of view the matrix
$A$ can be interpreted as the $N  \times N$ symmetric transition matrix of
the associated Markov process~\footnote{A random walk can be seen
as a Markov process with the identification position-state.}.

We are interested in the problem of the average MFPT on a node $s$ of
degree $k(s)$ of a random walker that started from a different,
unknown, node $x$. Our idea is mapping the original Markov
process $A$ on a much smaller process $B$ which will be asymmetric and
will contain self-loops (i.e. $B_{ii} \neq 0$). More precisely we
reduce the $N \times N$ matrix to a $O(\ln N / \ln \la k \ra \times \ln N /
\ln \la k \ra )$ matrix.  \\ Given the target node $s$, we start by
subdividing the entire network in subnetworks, or \ti{rings} (see
also~\cite{egocentered,Sood}), $r_l$, with the following property:

\be
r_l = \{nodes \;j\;|\;d_{sj}=l\}
\ee

\noindent where $d_{sj}=d_{js}$ is the distance between nodes $s$ and
$j$, i.e. the smallest number of links that a random walker has to
pass to get from $j$ to $s$. These rings will be the states of the new
matrix $B$. Their number, being proportional to the maximum distance
between any two nodes in the network, i.e. to the diameter of the
network, is $O(\ln N / \ln \la k \ra)$ \cite{chunglu1,chunglu2}
(where $ \la k \ra$ is the average degree of the nodes of the graph).
\\ Other important quantities are the average number $m_{r_l,r_{l+1}}
\equiv m_{l,l+1}$ of links that connect all the elements of $r_l$ with
all the elements of $r_{l+1}$, and the average number $m_{l,l}$ of
links between nodes belonging to the same $r_l$. We have trivially
$m_{l,l-1}=m_{l-1,l}$ and $m_{l,k}=0$ if $d_{lk}>1$.  \\ We now have
all the elements to define our new process. We are no more interested
in the exact position of the random walker. The relevant information
is now the ring in which the random walker is. The matrix of this
process has size $(l_{max}+1) \times (l_{max}+1)$, where $l_{max}$ is
the diameter of the original graph. The matrix has the following
structure~\footnote{For an easier notation rows (and columns) are
labeled with indexes starting from 0, instead of 1.} (for the
case $l_{max}=6$):

\be
 B = \left(
\begin{array}{c c c c c c c}
0 & 0  & 0  & 0  & 0  & 0  & 0 \\
b_{10} & b_{11}  & b_{12}  & 0  & 0  & 0  & 0 \\
0 & b_{21}   & b_{22}  & b_{23}  & 0  & 0 & 0 \\
0 & 0  & b_{32}  & b_{33}  & b_{34}  & 0 & 0 \\
0  & 0  & 0  & b_{43}  & b_{44}  & b_{45} & 0 \\
0  & 0  & 0  & 0  & b_{54}  & b_{55} & b_{56} \\
0  & 0  & 0  & 0  & 0 & b_{65}  & b_{66} \\
\end{array}
\right)
\label{eq:matB}
\ee \\ where $b_{ij} = m_{ij}/(\sum_{k \neq i \; =0}^{l_{max}}
m_{ik}+2m_{ii})$ for $i\neq j$, and $b_{ii} = 2m_{ii}/(\sum_{k \neq i
\; =0}^{l_{max}} m_{ik}+2m_{ii})$. $b_{ij}$ thus represents the
probability of going from ring $i$ to ring $j$. By definition of rings
it is clear that it is not possible to move from a ring to a
non-adjacent other ring, while it is obviously possible to move inside
a ring, and in this case the number of links must be doubled to take
into account that each internal link can be passed in two
directions. The elements of the first row of the matrix are set equal
to 0 because we are interested in the \ti{first} passage time in the
target node $s$. The probability $P_{ij}^{(t)}$ of going from state
$i$ to state $j$ in t steps is given by $(B^t)_{ij}$. If we set
$b_{01}=1$, we would allow the walker to escape from node $s$, while
$b_{00}=1$ should be used if we were interested in the probability
that the walker reached node $s$ before time $t$.  The probability
$F_{k(s)}(t)$ that the \ti{first} passage on node $s$ occurs at time
t is then: 

\be F_{k(s)}(t) = \sum_{l=1}^{l_{max}} \frac{n_l}{N-1} \;
{(B^t)_{l0}}
\label{eq:t} 
\ee

\noindent where $n_l$ is the number of nodes belonging to the ring
$r_l$ and each matrix term is weighted with the probability that the
random walker started in the ring corresponding to its row,
i.e. $n_l/(N-1)$.

\begin{figure}
\centerline{\psfig{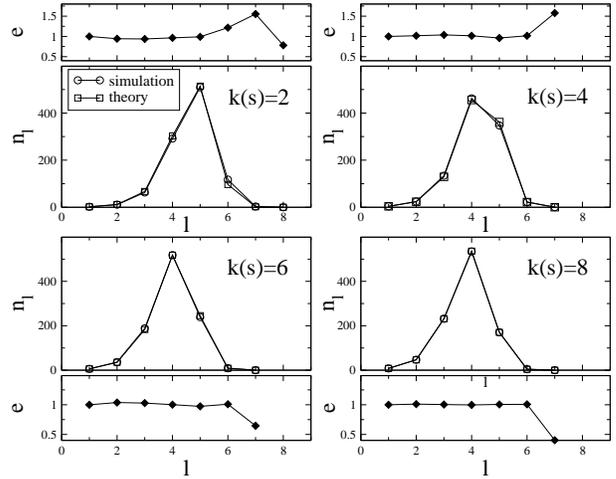}}
\vspace{0.2cm}
\caption{{\bf Nodes $n_l$ per ring:}  Rings populations $n_d$ are shown in
  this figure. Comparisons of theoretical previsions from
eq.(\ref{eq:n}) and data from simulations for different values of
$k(s)$ in a single E-R graph of size $N=10^3$ and $\la k \ra=6$ are
  shown. The fractional error $e$, defined as the ratio between measured and
  calculated values, is also plotted for each represented value of
  $k(s)$. Theoretical average quantities are in excellent agreement with
  single graph measurements.} 
\label{f:ring} 
\vspace{0.1cm} 
\end{figure} 

\noindent The average time MFPT $\tau(k(s))$ can be calculated using
eq.(\ref{eq:t}) as:

\be
\tau(k(s))) = \sum_{l=1}^{\infty} l F_{k(s)}(l) \label{eq:tau}
\ee

\begin{figure}
\centerline{\psfig{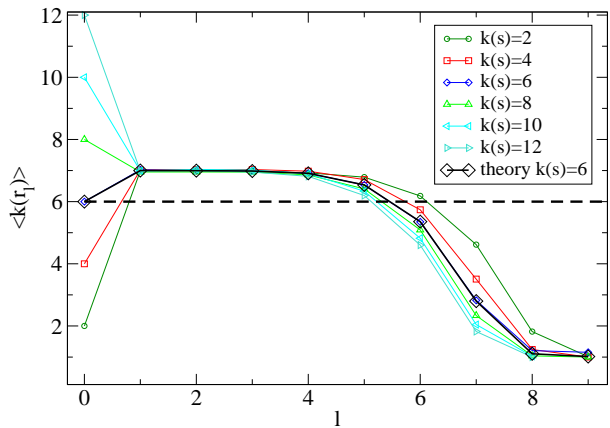}}
\caption{{\bf Average ring degree:} average degrees of nodes belonging
to different rings are shown in function of rings' distances for an ER
random graph ($N=10^4$ nodes and $\la k  \ra = 6$). A
dependence on ring's distance appears clearly, as predicted by
theory. Results for different values of $k(s)$ are shown (predictions
shown only for the case $k(s)=6$).  
} 
\label{f:gradoring} 
\vspace{0.1cm} 
\end{figure}
\section{Explicit calculation for random graphs}
\subsection{Static}
A random graph is obtained in the following manner: given a finite set
of isolated $N$ nodes, all the $N\times(N-1)/2$ pairs of nodes are
considered and a link between two nodes is added with probability
$p$. This yields (in the limit $N \rar \infty$) to Poisson's
distribution for the degree $k$ of a node:

\be
P(k)=\frac{\la k \ra ^k}{k!}e^{-\la k \ra} \label{eq:distpoisson}
\ee

\noindent with $\la k \ra = p(N-1)$. It is clear that such a graph
does not contain any relevant correlations between nodes and degrees,
and this will allow us to obtain exact average relations for the
quantities illustrated in the previous section.

The first important quantity to calculate is the average number $n_l$
of elements of $r_l$. It holds  
 
\be
n_{l+1} = \left(N - \sum_{k=0}^{l}n_k \right)(1 - (1-p)^{n_{l}}) \label{eq:n}
\ee

\noindent where $n_{l+1}$ is calculated as the expected number of nodes not
belonging to any interior ring that are connected with at least a member of
$r_l$. 
\noindent Figure~\ref{f:ring} illustrates that eq.(\ref{eq:n}) is in excellent
agreement with results from simulations. Obviously $n_0=1$ and
$n_1= \la k \ra$. However, if we know the degree $k(s)$ of node $s$ we
can impose $n_1=k(s)$ and calculate the following $n_{l>1}$ in the usual
manner. In fact, this is the way in which we will use eq.(\ref{eq:n}).

For the average numbers $m_{l,l+1}$ of links that connect all the
elements of $r_l$ with all the elements of $r_{l+1}$, and $m_{l,l}$ of
links between nodes belonging to the same $r_l$ we have: 

\bea
 m_{l,l+1} & = & n_l \left(N - \sum_{k=0}^{l}n_k
\right) p \label{eq:m} \\
m_{l,l} & = & \frac{n_l(n_l-1)}{2}p 
\nonumber
\eea

\noindent where the fact that a link between two nodes exists with
probability $p$ is exploited. As a practical prescription we add that when
eq.~(\ref{eq:m}) yields to non-physical $m_{l,l}<0$ one has to redefine
$m_{l,l}=0$. 
Also these expressions, which are crucial for the construction of the $B$
matrix of eq.~(\ref{eq:matB}), give predictions in excellent agreement with
results from simulations (data not shown). As expected we have that, in
the limit $p \rar 0$, $m_{l,l+1} \rar n_{l+1}$, while, when $p
\rar 1$, $m_{l,l+1} \rar n_l \times n_{l+1}$. Note that from eq.~(\ref{eq:m})
one has that $n_l<1 \; \Rightarrow \; m_{l,l} < 0$ which has clearly no
physical meaning.

Before going on it is interesting to make a remark. It is
known~\cite{egocentered,randomgen} that the nearest neighbors of a
generic node have particular properties, e.g. an average degree
different from the $\la k \ra$ of the graph.  Our relations are able
to predict this fact. Combining eq.(\ref{eq:n}) and eq.(\ref{eq:m}) it
is easy to see that the average degree of the nodes belonging to $r_l$
\ti{depends} on $l$, being constant (and larger than $\la k \ra$) for
low values of $l$ and decreasing rapidly for $l$ large enough. Data
from simulation reported in Figure~\ref{f:gradoring} show that this
prediction is correct. This agreement is not surprising since
eq.(\ref{eq:n}) and (\ref{eq:m}) are separately in excellent agreement
with simulations, thus being able to predict very accurately the value
of the average degree of the nodes belonging to the same ring.

\subsection{Dynamics}

\begin{figure}
\centerline{\psfig{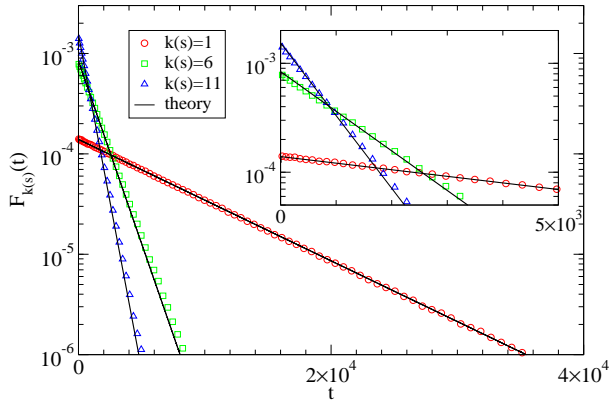}}
\caption{{\bf First passage time distributions:} FPT probability
distributions both measured and calculated for an ER graph with $\la k
\ra=6$ are presented for different values of $k(s)$. Theoretical
predictions are in excellent agreement with data of random walk on a
single graph. Theoretical curves are obtained with eq.(\ref{eq:t}),
and fit the relation $F_{k(s)}(t)=(1/\tau) exp(-t/\tau)$ with $\tau$ obtained
with eq.(\ref{eq:tau}). In the inset the first part of the distribution is
showed in more detail.}
\label{f:pfpt} 
\vspace{0.6cm} 
\end{figure} 

\begin{figure}[!h]
\centerline{\psfig{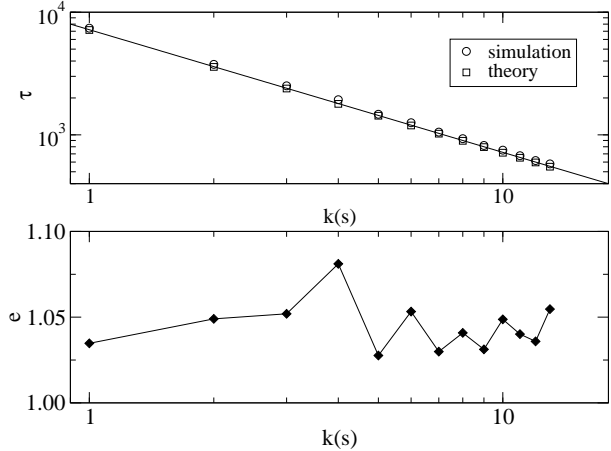}}
\caption{{\bf Mean first passage times:} in upper graph MFPT both measured and
calculated (using eq.(\ref{eq:tau})) are reported for an ER graph of size
$N=10^3$ with $\la k \ra=6$. Error bars on measured values are not visible on
the scale of the graph. The line $\tau(k(s)) \simeq \tau(1) \times k(s)^{-1}$
is also 
plotted. It holds $\tau(1)_{sim}=7413$ and $\tau(1)_{calc}=7164$ for values
obtained respectively from simulations and from calculation. It can be noted
that the order of magnitude of $\tau(1)$ is given by $2M$, where $M$ is the
total number of links in the graph; in our case we have $2M \simeq \la k \ra
\times N = 6000$.  In lower graph the fractional error $e$, defined as the
ratio between simulated and calculated MFPTs, is reported.}
\label{f:meantime} 
\vspace{0.1cm} 
\end{figure} 

So far, we have shown that predictions of the above relations on the
 static properties of a random graph are correct. We now explore the
 predictions on the diffusion processes.  Figure~\ref{f:pfpt} shows,
 for different values of the degree $k(s)$ of the target node, the
 comparison of the predicted values of ${F}_{k(s)}(t)$, calculated
 with eq.(\ref{eq:t}) and results from simulations. Data from
 simulations are obtained by selecting randomly, for each of the
 approximately $N \times N$ runs, one of the nodes of selected degree
 $k(s)$ present in the network to play the role of $s$, and one of the
 remaining $N-1$ nodes to be the starting point of the random walker.
 The agreement between theory and simulations is very good. The
 exponential behavior, typical of finite ergodic Markov chains, has
 the form $f(t)=(1/\tau) exp(-t/\tau)$.

Figure~\ref{f:meantime} shows that the average MFPT $\tau(k(s))$,
 calculated using eq.(\ref{eq:tau}) (with the $B$ matrix built using
 (eq.(\ref{eq:n}) and (\ref{eq:m})), are in good agreement (though
 slightly smaller) with those obtained in simulations. We shall return
 to the origin of the small disagreement between theory and
 simulations at the beginning of the next section.

The relation, found both in theory and simulations (see
 also~\cite{nohrieger}), $\tau(k(s)) = \tau(1)\times k(s)^{-1}$, can
 be explained with elementary qualitative probabilistic arguments. In
 fact, since, as shown in Figure~\ref{f:gradoring}, the average degree
 of the nodes in $r_1$ does not depend on $k(s)$ (i.e. on the size
 $n_1$ of $r_1$), also the MFPT on a node of $r_1$ is independent from
 $n_1$. This means that, while the probability of passing from $r_1$
 to $s$ does not depend on $n_1$, a larger $r_1$ is visited more often
 than a smaller one. Combining these observations, it seems plausible
 guessing that the MFPT on a target node $s$ with $k(s)>1$ will be
 $1/k$ times the MFPT of a node $s$ with $k=1$, and this behavior is
 indeed observed.

\begin{figure}[t]
\centerline{\psfig{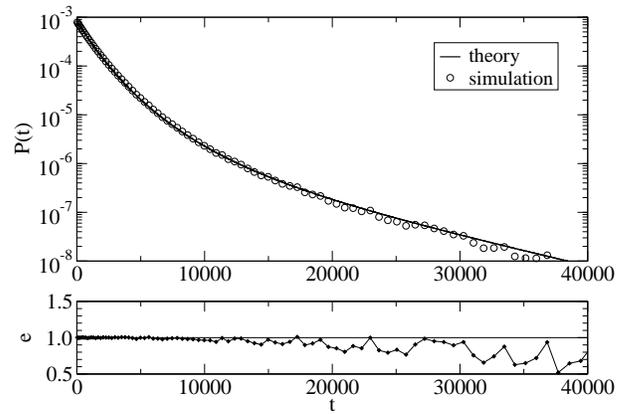}}
\caption{{\bf First passage time distribution - Average on all target
nodes $s$:} The distribution of the FPT obtained is shown (ER graph:
$N=10^3$, $\la k \ra=6$). At each run both the target node $s$ and the
starting node of the random walker are randomly chosen. This means
that in this distribution the degree of the target nodes is no more
fixed. The non exponential curve is the result of the convolution of
several exponential curves obtained for fixed $k(s)$. The theoretical
curve, obtained according to eq.(\ref{eq:T}), is in excellent
agreement with data from simulation performed on a single graph. The
fractional error $e$, defined as the ratio between simulated and calculated
data, is also plotted. For large times poor statistics causes
bigger fluctuations of $e$.}
\label{f:allgoal} 
\vspace{0.9cm} 
\end{figure} 

It is worth noting that both the $k(s)^{-1}$ trend and the order of magnitude
  of $\tau(1)$ can be derived with a simple mean field
  approach~\cite{Saramaki, Evans}. In fact, once neglected all the
  possible correlations in a graph, the whole random walk process can be
  approximated with a two state Markov process where the two states correspond
  to the walker being at the target node and on any other node. Easy
  calculations shows that the probability for the walker to arrive at a node
  $s$ is given by $q(s)=q(k(s))=k(s)/2M$, where $M$ is the total number of
  links of the graph. For a fully connected graph this relation gives the
  exact value of $\tau=\tau(N-1)=N-1$. In a random graph the mean field
  approach gives better and better results the larger is the mean degree $\la
  k \ra$. For small values of $\la k \ra$, only the order of magnitude of
  $\tau(1)$ is in fact predicted by this approach. The method based on rings,
  thought being less simple, is able to make more accurate predictions for all
  values of $\la k \ra$. Just for comparison we report here data shown in
  Figure~\ref{f:meantime} relative to a network of $N=10^3$ nodes with $\la k
  \ra=6$ : we have $\tau(1)_{sim} / \tau(1)_{calc} = 1.03$ and $\tau(1)_{sim}
  / 2M = 1.24$, where $ \tau(1)_{sim}$ and $\tau(1)_{calc}$ are MFPT obtained
  respectively with simulations and with the ring method.

All the results discussed above allow us to explain the curve presented in
Figure~\ref{f:allgoal}, which represents the distribution $P(t)$ of the MFPT
in a graph when both $s$ and the starting point of the walker are randomly
chosen at each run. $P(t)$ can be calculated here as the convolution of
several exponential FPT distributions $F_{k(s)}(t)$ corresponding to the
different values of $k(s)$, each weighted with the probability of encountering
a node of degree $k(s)$ in the graph. More precisely, according to
eq.(\ref{eq:distpoisson}), for each time step $t$ every $F_{k(s)}(t)$ must be
weighted with Poisson's weights $c_{k(s)}^{(\la k \ra)}=\frac{\la k
\ra^{k(s)}}{k(s)!}e^{-\la k \ra}$. We have:

\bea P(t) &=& \sum_{j=1}^{\infty} c_{j}^{(\la k \ra)} F_j(t) \nonumber \\ &=&
\sum_{j=1}^{\infty} c_{j}^{(\la k \ra)} e^{(-t/\tau(j))} \frac{1}{\tau(j)}
\label{eq:T} 
\eea

\noindent This relation can be written in a more compact way exploiting the
fact that $\tau(k(s)) = \tau(1)\times k(s)^{-1}$~\cite{dalla}. Defined $Z=\la
k \ra \times exp(-t/\tau(1))$, it holds:

\bea
P(t) &=& \sum_{j=1}^{\infty} Z \frac{Z^{j-1}}{(j-1)!}  
\frac{e^{-\la k \ra}}{\tau(1)}= c Z e^Z
\eea

\noindent where $c$ is the constant $\frac{e^{-\la k \ra}}{\tau(1)}.$

\section{Extensions of the theory}

In the previous sections we have described a method that allows to
calculate the average MFPT on a node $s$ of a walker that started from
a generic other node of the graph. We have then obtained exact
(average) expressions for the case of random graphs. Unfortunately,
the analytical extension of the relations found for this kind of
graphs to other graphs (such as, for example, scale free networks) is
difficult. This is due to the fact that eq.(\ref{eq:n}) and
eq.(\ref{eq:m}) exploit the knowledge of the rules according to which
a random graph is generated. In other words the absence of
correlations between nodes is the main feature those equations are
based on. When correlations are present the calculation of the number
of nodes of the second ring, $n_2$, is already very difficult (for
finite networks) and requires some empirical
assumptions~\cite{randomgen}.

\begin{figure}
\centerline{\psfig{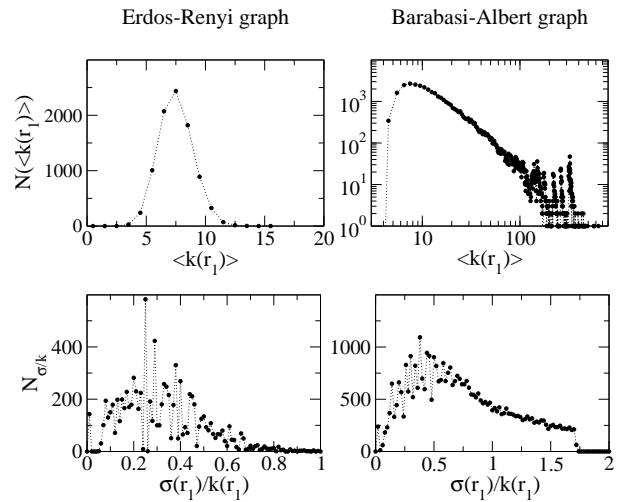}}
\caption{{\bf First ring fluctuations:} Fluctuations related to nodes
   of degree $k=3$ are presented for ER (left) and BA (right) graphs
   of size $N=1 \times 10^5$ and $\la k \ra =6$. On the two top
   figures global fluctuations are analyzed. Histograms of the number
   of nodes vs. the average degree $k(r_1)$ of the nodes of the first
   ring $r_1$ for ER graphs (top-left) and BA graphs
   (top-right). Logarithmic scale for the ordinate of BA graph must be
   noted. In the two bottom figures local fluctuations are
   analyzed. Here the histograms represent the number of rings in
   function of the ratio $\sigma (r_1)/k(r_1)$, where $\sigma(r_1)$ is
   the variance of the degree of the nodes belonging to each $r_1$ for
   ER graphs (top-left) and BA graphs (top-right). It is evident the
   higher degree of fluctuations of the BA graphs.}
\label{f:dis} 
\vspace{0.9cm} 
\end{figure}
  
In addition there is a more subtle reason that makes our method
difficult to extend. Given the set of all nodes of a graph with a
certain degree $k$, their first rings, although having the same number
of nodes, present two kinds of fluctuations. On a global scale, the
average degree $k(r_1)$ of the nodes of the first ring does not have a
unique value, but in general is distributed according to some
probability density. On a local scale, on the other hand, a single
ring is not made by identical nodes, and its average degree has a
certain variance $\sigma$. In Figure~\ref{f:dis} we show global and
local fluctuations for both a random graph and a Barabasi-Albert (BA)
scale free graph~\cite{barabasiI}. The preferential attachment rule of
the Barabasi-Albert network generates a graph with a scale free form
$P(k)\simeq k^{-c}$, with $c=3$, for the degree distribution. As it is
evident from Figure~\ref{f:dis}, BA graphs have larger fluctuations
than random graphs.

With our method of rings, described before, the fluctuations cannot be
taken into account. Matrix $B$ (\ref{eq:matB}) is in fact defined
under both the assumptions of (i) equivalence of all the nodes with a
given $k$ (global homogeneity) and (ii) equivalence of all the nodes
inside each ring (local homogeneity). The slight disagreement between
our theory and simulation results present in Figure~\ref{f:meantime}
are thus easily explained in term of local fluctuations of the first
ring. In fact, as it is easy to demonstrate using Lagrange
multipliers, the assumed local ordered configuration is the most
advantageous for a walker that has to reach the node $s$ from the
first ring. This is then the reason for which our calculated MFPT are
always smaller than those obtained from simulations.

\begin{figure}
\centerline{\psfig{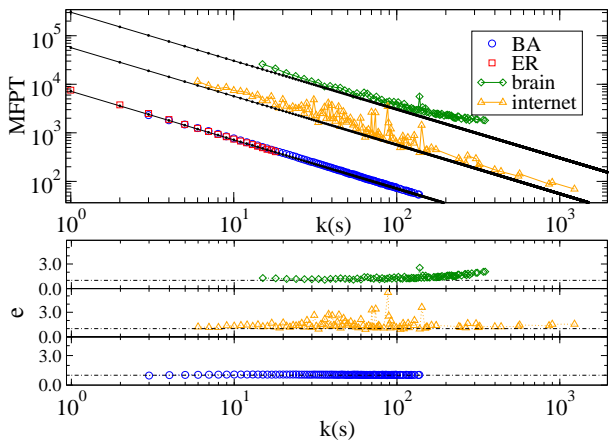}}
\caption{{\bf Mean first passage times vs. the degree of the target
  node for different networks.} Continuous lines with small filled points
  (upper graph) are obtained from eq.(\ref{eq:tau}), i.e. for random ER
  graphs, while empty 
  symbols of different shapes come from simulations. It is evident the
  excellent agreement between theory and simulations both for ER graphs
  (circles) and, less obviously, for BA graphs (squares). Data from
  real networks are also reported: also in this case the agreement
  with theoretical predictions is good. Error bars are not visible on the
  scale of the graph. In lower graph fractional errors $e$, defined as the
  ratio between measured and calculated values, are reported
  both for real networks and BA graph. Dashed lines indicate correspond to
  $e=1$. }
\label{f:mfpt} 
\vspace{0.9cm} 
\end{figure}

Notwithstanding these difficulties in extending our theory, we found a
quite surprisingly result, shown in Figure~\ref{f:mfpt}: given a BA
graph with a given average degree $\la k \ra$, the average MFPT for a
walker starting from a generic node on a node $s$ of degree $k(s)$ is
almost equal to the corresponding average MFPT of the same random walk
on a random graph with the same average degree. This means that our
theory continues to predict very well the MFPT (and hence its
exponential distribution). It is remarkable that the theory predicts
well also the MFPT on nodes with high degree, which are absent in the
corresponding random graph.

The ability of our theory to predict diffusion processes on BA graphs
can be due to the modest presence  of correlations between its
nodes. Many properties of real networks, in fact, are not reproduced
by the BA model. One important measure of correlations in a graph
is the measure of the average degree of the nearest neighbors (i.e. of
the nodes of the first ring) of vertices of degree $k$, called
$k_{nn}$ \cite{knn}. While random and BA graphs have a flat $k_{nn}$,
indicating the absence of strong correlations among nodes, many real networks
exhibit either an assortative or disassortative behavior. In the
first case high degree vertices tend to be neighbors to high degree
vertices, while in the second case they have a majority of low degree
neighbors. Another important measure of correlation is the clustering
coefficient which is proportional to the probability that two neighbors
of a given node are also neighbor of themselves. Again, BA and random graphs,
in which clustering is very poor, do not reproduce the
clustering properties of many real networks.

In order to check how far our theory can predict the MFPT on
correlated graphs we have performed two sets of experiments on real
networks. We have considered in particular a network of Internet at
the level of Autonomous Systems~\cite{cosin,ale_libro} which exhibits a
disassortative mixing feature and a recently proposed scale-free brain
functional network~\cite{brain} which exhibits an assortative mixing
feature as well as a strong clustering coefficient.

The results for the MFPT for these two networks, as a function of the
degree of the target node, are reported in Figure~\ref{f:mfpt}.
Though the agreement between theory and simulation is not any more
perfect, it remains good. In particular we find again the approximate
trend $\tau(k(s)) = \tau(1)\times k(s)^{-1}$.

\begin{figure}
\centerline{\psfig{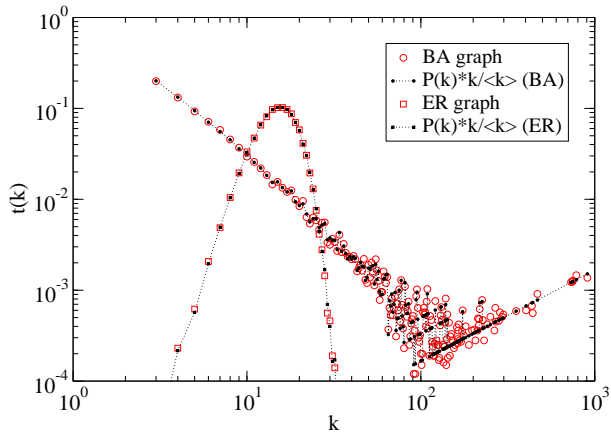}}
\caption{{\bf Degree distribution exploration:} a random walker
  explores the degree distribution of two networks: a BA graph of size
  $N=10^5$ and $\la k \ra=6$ and an ER random graph of size $N=10^5$
  and $\la k \ra=15$. Simulations in which a walker travels the
  networks for $N$ time steps have been performed. Empty symbols in figure
  represent the fraction of time spent on a node of degree $k$. Filled symbols
  joined by light lines are obtained with the relation $P(k)k/ \la k \ra$,
  where $P(k)$ is the degree distribution of the considered
  network. Filled points fit well the experimental data, and, as it is
  obvious, a longer walking process would allow for better
  agreement. The probability of finding a random walker on a node of a
  given degree $k$ is related to the degree distribution of the graph
  via the inverse of $1/k$ MFPT scaling relation (apart from a
  normalizing factor). For BA graphs it holds $P(k)\propto
  k^{-3}$. The first part of the curve $P(k)k/ \la k \ra$ presented in
  Figure is indeed $c \times k^{-2}$, where $c$ is a constant. The
  region of higher degrees, on the other hand, grows linearly due to
  the fact that in a finite size realization the statistics on high
  degree nodes is poor and the degree distribution in this region is
  flat. We avoided any binning in experimental or theoretical data to make
  clear that the walker exploration of the network is not perfect. }
\label{f:degreeexpl} 
\vspace{0.9cm} 
\end{figure} 

Now we have all the elements to estimate the probability of finding
a random walker on a node of a given degree $k$. On the one hand, in
fact, it seems obvious that this probability is related to the
fraction $f(k)$ of nodes of that degree in the network, while on the
other hand we now know that the MFPT on such a node is proportional,
on average, to $1/k$. It is then reasonable arguing that the
probability for a random walker being on a node of degree $k$ is
proportional to $kf(k)$.

\noindent We have tested this hypothesis in an experiment reported in
Figure~\ref{f:degreeexpl}. In the experiment a walker has explored a
BA network and an ER random graph with $N=10^5$ nodes for $N$ time
steps. At each time step the degree of the visited node was recorded
and the normalized histogram of the fraction of time spent on nodes of
any degree is reported in Figure~\ref{f:degreeexpl}. In the limit of
infinite time steps this histogram would indicate exactly the
probability of finding the walker of a node of a given $k$. According
to our hypothesis, this histogram should be described by the function
$P(k)k/\la k \ra$, where $P(k)$ is the degree distribution of the
considered network, and the Figure shows this is in fact the case,
already after a relatively small number of time steps.

\noindent Finally, it is worth noting that the previous argument can
be reversed. A walker able to record the degree of each node it
traverses can be used to determine the degree distribution of the
network it travels. In fact, if $t(k)$ is the fraction of time spent
on nodes of degree $k$ it holds $f(k) \propto t(k)/k$. The average
degree $\la k \ra$ is then trivially obtained by requiring the
normalization of the estimated $P(k)$.

\section{Conclusions}

In this paper we addressed the problem of the computation of the mean first
passage time on a selected node $s$ of random walkers starting from different
nodes on a generic network. We have introduced a new approximate method, based
on the concept of rings, which maps the original Markov process on another
Markov process in a much smaller space. This allows for a drastic reduction of
the computational cost.  In the case of ER random graphs we have been able to
analytically derive all the quantities of interest and we have shown that our
method gives predictions, both for static and dynamic properties, in excellent
agreement with results found in simulations. Even if this new method is
promising, analytical results are difficult to obtain for non random
graphs. However, quite surprisingly, we have found that MFPT calculated with
our theory for ER graphs are in excellent agreement also with simulations of
dynamics on BA networks and in good agreement with results obtained with
random walkers on two real networks, thus making our method an easy tool to
predict MFPT time related quantities in many cases. Acknowledgments: This
research has been partly supported by the ECAgents project funded by the
Future and Emerging Technologies program (IST-FET) of the European Commission
under the EU RD contract IST-1940. The information provided is the sole
responsibility of the authors and does not reflect the Commission's
opinion. The Commission is not responsible for any use that may be made of
data appearing in this publication. Brain data collection was supported with
funding from NIH NINDS of USA (Grants 42660 and 35115). The authors are
grateful to Claudio Castellano and Alessandro Taloni for many interesting
discussions and Alain Barrat and Luca Dall'Asta for a careful reading of the
manuscript.

\end{document}